\documentstyle[12pt]{article}
\catcode `\@=11
\@addtoreset{equation}{section}

\def\qed{\vrule height 6pt width 6pt depth 6pt}

\newcommand{\be}{\begin{equation}}
\newcommand{\ee}{\end{equation}}
\newcommand{\R}{\rm I \mkern -3mu R}
\newcommand{\N}{\rm I \mkern -3mu N}
\newtheorem{thm}{Theorem}
\newtheorem{rem}{Remark}
\newtheorem{lemma}{Lemma}
\newtheorem{cor}{Corollary}
\newtheorem{prop}{Proposition}
\begin{document}
%\thispagestyle{empty}
%\begin{flushright}
%\end{flushright}
%\bigskip\bigskip
\begin{center}
{\bf \Large{TRIVIAL SECOND-ORDER LAGRANGIANS IN ~\\
CLASSICAL FIELD THEORY}}
\end{center}
\vskip 1.0truecm
\centerline{D. R. Grigore
\footnote{e-mail: grigore@theor1.ifa.ro, grigore@roifa.ifa.ro}}
\vskip5mm
\centerline{Dept. of Theor. Phys., Inst. Atomic Phys.}
\centerline{Bucharest-M\u agurele, P. O. Box MG 6, ROM\^ANIA}
\vskip 2cm
\bigskip \nopagebreak
\begin{abstract}
\noindent
Trivial second-order Lagrangians are studied and a complete description of the
dependence on the second-order derivatives is given. This extends previous work
of Olver and others. In particular, this description involves some polynomial 
expressions called hyper-Jacobians. There exists some linear dependencies 
between these polynomials which are elucidated for the (second-order) 
hyper-Jacobians.
\end{abstract}
%\newpage\setcounter{page}1

\section{Introduction}

The problem of classification of trivial Lagrangians i.e. Lagrangians 
leading to trivial Euler-Lagrange equations has a long history and it seems
that it is not solved in full generality even today in the framework of
classical field theory. The basic input to the rigorous formulation of this
problem is a fiber bundle
$
\pi: Y \rightarrow X
$
where
$X$
is usually interpreted as the space-time manifold and the fibres describe the
internal degrees of freedom of the problem. To formulate the Lagrangian
formalism one has to include the ``velocities" i.e. one must work on some
jet-bundle extension of
$
\pi: Y \rightarrow X.
$

There are two approaches to this problem. The first one uses the infinite jet
bundle extension formalism. In this framework one can prove that, quite
generally, a trivial Lagrangian is what is usually called, a total derivative 
\cite{T}, \cite{A}. The second approach, which is more suitable for the needs
of specific physical problems is to use a finite jet bundle extension. This
corresponds to field theories with a finite order of partial derivatives. The
most common physical theories are, at most, of second order. The r\^ole of the
finite-order approach has been particulary emphasised by Krupka (see for
instance \cite{K1}). In the finite-order approach there is no general solution 
of the trivial Lagrangian problem. However, there are a number of partial 
results which deserve mentioning. For instance, the first-order case has 
been completely elucidated (see \cite{E}-\cite{GP}; needless to say, this 
rather simple case has been rediscovered independently by many people from 
time to time). For the second and higher order Lagrangians only partial 
results are known to the author. More precisely, we refer to Olver's papers 
\cite{BCO}, \cite{O} where one studies the case of a trivial Lagrangian of 
order 
$r$
which is dependent only of the highest order derivatives in a polynomially 
homogeneous way. In this case one can show that the Lagrangian is a linear
combination of some polynomial expressions called hyper-Jacobians. The
existence of some linear dependence between the hyper-Jacobians is emphasised
but the general problem of finding all possible linear dependencies is not
solved. It is worth mentioning that the results of Olver are based on a very
complex algebraic machinery of higher dimensional determinants, invariant
theory, etc. combined with a generalization of Gel'fand-Dikii transform. 

In this paper we will present a rather complete analysis of case of the 
second-order trivial Lagrangian. Without any limitations we will be able to 
prove that the dependence of the second order derivatives is through the 
hyper-Jacobian polynomials. In particular the polynomiality emerges naturally 
from the proof and the homogeneity condition is not needed. Moreover we will 
elucidate the  question of possible linear dependencies between the 
hyper-Jacobians and it  will emerge that only the tracelessness conditions 
already appearing in  \cite{BCO}, \cite{O} are sufficient i.e. no other 
constraints are independent of these. All these results will be obtained 
from a long but quite elementary proof. The methods used are essentially 
those used in \cite{G1}, \cite{G2} for the analysis of the most general 
form of a Euler-Lagrange expression: they consist in complete induction and 
some Fock space techniques. 

We feel that this completely new method may offer a way for the analysis of 
the general case of
$r$-th
order Lagrangians and so it deserves some attention.

The structure of the paper is the following one. In Section 2 we present the
jet-bundle formalism to fix the notations and illustrate our problem on the
well-known first-order trivial Lagrangian case. We also present the equations
to be solved in the second-order case and comment on the best strategy to solve
them. In Section 3 we consider the second-order case for a scalar field theory.
The proof is based on induction on the dimension of the space-time manifold and
is contained in the proof of the main theorems in \cite{G1}, \cite{G2}. We give
it, nevertheless in this Section because we will need further some notations,
results and techniques. We also start the analysis for the general case by an
obvious corollary of the result obtained for the scalar field and we will
formulate the general result we want to prove. In Section 4  the case of a 
two component classical field is analysed and one is able to  perceive the 
nature of the difficulties involved in this analysis. In  Section 5 the 
case of a field with three or more components is settled  completing the 
analysis. In Section 6 we combine the results from the preceding two 
Sections in the main theorem and make some final comments. 

\section{The Jet-Bundle Formalism in the Lagrangian Theory}

2.1 As we have said in the Introduction, the kinematical structure of 
classical field theory is based on a fibred bundle structure
$
\pi: Y \mapsto X
$
where
$Y$
and
$X$
are differentiable manifolds of dimensions
$
dim(X) = n, \quad dim(Y) = N + n
$
and
$\pi$
is the canonical projection of the fibration. Usually $X$
is interpreted as the ``space-time" manifold and the fibres of $Y$
as the field variables. Next, one considers the $r$-th jet bundle
$
J^{r}_{n}(Y) \mapsto X \quad (r \in \N).
$
A 
$r$-th
order jet with source
$x \in X$,
and target
$y \in Y$
is, by definition, an equivalence class of all the smooth maps
$
\zeta: X \rightarrow Y
$
verifying
$\zeta(x) = y$
and having the same partial derivatives in $x$ up to order $r$ (in any
chart on $X$ and respectively on $Y$).
We denote the equivalence class of
$\zeta$
by
$
j^{r}\zeta
$
and the factor set by
$
J^{r}_{x,y}.
$
Then the
$r$-th order
jet bundle extension is, by definition
$
J^{r}_{n}(Y) \equiv \cup J^{r}_{x,y}.
$

One usually must take
$
r \in \N
$
sufficiently large such that all formulas make sense. Let us consider a
local system of coordinates in the chart
$
U \subseteq X: \quad
(x^{\mu}) \quad (\mu = 1,...,n).
$
Then on some chart
$
V \subseteq \pi^{-1}(U) \subset Y
$
we take a local coordinate system adapted to the fibration structure: 
$
(x^{\mu},\psi^{A}) \quad (\mu = 1,...,n, \quad A = 1,...,N) 
$
such that the canonical projection is
$
\pi(x^{\mu},\psi^{A}) = (x^{\mu}).
$

Then one can extend this system of coordinates to
$
J^{r}_{n}(Y)
$
as follows: on the open set
$
V^{r} \equiv (\pi^{r,0})^{-1}(V)
$
we define the coordinates of
$
j^{r}_{x}\zeta
$
to be
$
(x^{\mu},\psi^{A},\psi^{A}_{\mu},...,\psi^{A}_{\mu_{1}.,,,,\mu_{r}})
$
where
$
\mu_{1} \leq \cdots \leq \mu_{s} \qquad (s \leq r).
$
Explicitly

\be
\psi^{A}_{\mu_{1},...,\mu_{s}}(j^{r}_{x}\zeta) \equiv
\prod_{i=1}^{s} {\partial\over \partial x^{\mu_{i}}} \zeta(x) 
\quad (s=1,...,r).
\ee

If
$
\mu_{1},...,\mu_{s}
$
are arbitrary numbers belonging to the set
$
\{1,...,n\}
$
then by the expression
$
\{\mu_{1},...,\mu_{s}\}
$
we understand the result of the operation of increasing ordering. Then
the notation
$
\psi^{A}_{\{\mu_{1},...,\mu_{s}\}}
$
becomes meaningful for all set of numbers
$
\mu_{1},...,\mu_{s}.
$
If
$
I = \{\mu_{1},...,\mu_{s}\}
$
is an arbitrary set from
$
\{1,...,n\}^{\times s}
$
then we define

\be
\psi^{A}_{I} = \psi^{A}_{\mu_{1},...,\mu_{s}} \equiv
\psi^{A}_{\{\mu_{1},...,\mu_{s}\}}.
\ee

This notation makes sense whenever the cardinal of $I$ verifies:
$
|I| \leq r
$
where if
$
I = \emptyset
$
then we put
$
\psi^{A}_{\emptyset} = \psi^{A}.
$
With this convention the expression
$
\psi^{A}_{I}
$
is completely symmetric in the individual
$
\mu_{1},...,\mu_{s}
$
which make up the multi-index $I$.

2.2 Let us consider
$
s \leq r
$
and
$T$
a
$(n + 1)$-form
which can be written in the local coordinates introduced above as:

\be
T = {\cal T}_{A} \quad d\psi^{A} \wedge dx^{1} \wedge \cdots \wedge dx^{n}
\label{edif}
\ee
with
$
{\cal T}_{A}
$
some smooth functions of
$
(x^{\mu},\psi^{A}_{I}) \qquad (|I| \leq s).
$

Then
$T$
is a globally defined object. We call such a
$T$
a {\it differential equation of order s}.

2.3 To introduce some special type of differential equations we need some
very useful notations \cite{AD}. We define the differential operators:
\be
\partial^{I}_{A} \equiv {r_{1}!...r_{l}! \over |I|!}
{\partial \over \partial \psi^{A}_{I}}
\label{pdif}
\ee
where
$
r_{i}
$
is the number of times the index
$i$
appears in $I$.
The combinatorial factor in (\ref{pdif}) avoids possible overcounting in the
computations which will appear in the following. One has then:

\be
\partial^{\mu_{1},...,\mu_{l}}_{A} \psi^{B}_{\nu_{1},...,\nu_{m}} =
\cases{  { 1\over l!} \delta^{A}_{B}
perm(\delta^{\mu_{i}}_{\nu_{j}}), & for $l = m$ \cr
0,  & for $l \not= m$ \cr}
\ee
where

\be
perm\left( \delta^{\mu_{i}}_{\nu_{j}} \right) \equiv
\sum_{P \in {\cal P}_{l}} \delta^{\mu_{1}}_{\nu_{P(1)}}\cdots
\delta^{\mu_{l}}_{\nu_{P(l)}}
\ee
is a permanent. (In general we denote by
$
perm(A)
$
the permanent of the matrix
$A$).

Next, we define the total derivative operators:

\be
D_{\mu} = {\partial\over \partial x^{\mu}} + \sum_{l=0}^{r-1}
\psi^{A}_{\nu_{1},...,\nu_{l},\mu} \partial^{\nu_{1},...,\nu_{l}}_{A}
= {\partial\over \partial x^{\mu}} + \sum_{|I|\leq r-1} \psi^{A}_{I\mu}~
\partial^{I}_{A}
\label{tdif}
\ee
where we use the convention
$
IJ \equiv I \cup J.
$
One can check that

\be
D_{\mu}\psi^{A}_{I} = \psi^{A}_{I\mu}, \qquad |I| \leq r-1
\label{der}
\ee
and
\be
[D_{\mu}, D_{\nu}] = 0.
\label{com}
\ee

Finally we define the differential operators

\be
D_{I} \equiv \prod_{i \in I} D_{\mu_{i}}.
\label{tdifs}
\ee

Because of (\ref{com}) the order of the factors in the right hand side is
irrelevant.

2.4 A differential equation
$T$
is called {\it locally variational} (or of the {\it Euler-Lagrange type})
{\it iff} there exists a local real function
${\cal L}$
such that the functions
$
{\cal T}_{A}
$
from (\ref{edif}) are of the form:

\be
{\cal E}_{A}({\cal L}) \equiv \sum_{l=0}^{r} (-1)^{l}
D_{\mu_{1},...,\mu_{l}} (\partial^{\mu_{1},...,\mu_{l}}_{A} {\cal L})
\label{Eop}
\ee

One calls
${\cal L}$
a {\it local Lagrangian} and:

\be
L \equiv {\cal L}~dx^{1}\wedge\cdots \wedge dx^{n}
\label{Lform}
\ee
a {\it local Lagrange form}.

If the differential equation
$T$
is constructed as above then we denote it by
$
E(L).
$
A local Lagrangian is called a {\it total divergence} if it is of the form:

\be
{\cal L} = D_{\mu} V^{\mu}.
\ee

A Lagrangian is called {\it trivial} (or {\it null} in the terminology of
\cite{BCO}, \cite{O}) if it satisfies:

\be
E(L) = 0.
\label{trEL}
\ee

One can check that a total divergence Lagrangian is trivial. The converse of 
this statement has been proved only in the infinite jet bundle approach
\cite{T}.  

2.5 Let us briefly review the case of trivial first-order Lagrangians.
One must take in the equation (\ref{trEL}) above the function
${\cal L}$
depending only on the variables
$
(x^{\mu}, \psi^{A}, \psi^{A}_{\mu}).
$

Then we obtain the condition of triviality as follows:

\be
\partial_{A} {\cal L} - D_{\mu} \partial_{A}^{\mu} {\cal L} = 0.  
\label{trEL1}
\ee

One can easily find out that this equation is equivalent to the following 
two equations:

\be
\left(\partial^{\mu}_{A}~\partial^{\nu}_{B} + \mu \leftrightarrow \nu
\right) {\cal L} = 0
\label{I1}
\ee
and

\be
\partial_{A} {\cal L} - 
(\partial_{\mu} + \psi^{B}_{\mu} \partial_{B})
\partial_{A}^{\mu} {\cal L} + 
(\partial_{\mu} + \psi^{B}_{\mu} \partial_{B}) 
(\partial_{\nu} + \psi^{C}_{\nu} \partial_{C}) {\cal L} = 0.
\label{I2}
\ee

From (\ref{I1}) one easily discovers that
${\cal L}$
is a polynomial of maximal degree
$n$
in the first-order derivatives of the following form:

\be
{\cal L} = \sum_{k=1}^{n} {1 \over k!} 
L^{\mu_{1},...,\mu_{k}}_{A_{1},...,A_{k}}
\prod_{i=1}^{k} \psi^{A_{i}}_{\mu_{i}}
\label{TrI}
\ee
where the functions
$
L^{...}_{...}
$
depend only on the variables
$
(x^{\mu}, \psi^{A})
$
and are completely antisymmetric in the upper indices
$
\mu_{1},...,\mu_{k}
$
and in the lower indices
$
A_{1},...,A_{k}.
$

To exploit the equation (\ref{I2}) one defines the form

\be
\Lambda = \varepsilon_{\mu_{1},...,\mu_{n}} \quad
\sum_{k=1}^{n} {1 \over k1} L^{\mu_{1},...,\mu_{k}}_{A_{1},...,A_{k}} \quad
d\psi^{A_{1}} \wedge \cdots \wedge d\psi^{A_{k}} \wedge
dx^{\mu_{k+1}} \wedge \cdots \wedge dx^{\mu_{n}}
\label{lam}
\ee
and shows that (\ref{I2}) is equivalent to
\be
d\Lambda = 0.
\ee

2.6 Finally we give the similar details for the second-order case. 
Let us consider a second-order Lagrangian:
$
{\cal L}(x^{\mu} ,\psi^{A},\psi_{\nu}^{A},\psi_{\nu \rho }^{A}) 
$
and impose the triviality condition (\ref{trEL}). One obtains explicitly:

\be
\partial_{A} {\cal L} - D_{\mu} \partial_{A}^{\mu} {\cal L} + 
D_{\mu} D_{\nu} \partial_{A}^{\mu\nu} {\cal L} = 0.  
\label{trEL2}
\ee
We write in detail this equation and note a linear dependence on the
fourth-order derivatives; applying the operator
$
\partial^{\zeta_{1}\zeta_{2}\zeta_{3}\zeta_{4}}_{B}
$
we obtain after some rearrangements:

\be
\sum_{(\rho_{1},\rho_{2},\rho_{3})}
\partial_{A}^{\mu\rho_{1}} \partial_{B}^{\rho_{2}\rho_{3}}  
{\cal L} + (A \leftrightarrow B) = 0.
\label{II1}
\ee

We will use many times from now on a convenient notation, namely by
$
\sum_{(\mu,\nu,\rho)}
$
we will mean the sum over all the {\it cyclic} permutations of the indices
$
\mu, \nu, \rho.
$

We take into account this equations in the original equation (\ref{trEL2}) 
to simplify it a little bit by eliminating the dependence on the fourth order
derivatives. What remains is an equations having a quadratic dependence on the
third-order derivatives. We differentiate twice this equation with respect to 
the third-order derivatives and obtain as before:
 
\be
\sum_{(\rho_{1},\rho_{2},\rho_{3})} \sum_{(\zeta_{1},\zeta_{2},\zeta_{3})} 
\partial^{\zeta_{1}\zeta_{2}}_{D}\partial^{\zeta_{3}\rho_{3}}_{A}
\partial^{\rho_{1}\rho_{2}}_{B} {\cal L} = 0.
\label{II2}
\ee

Taking into account (\ref{II1}) and (\ref{II2}) the initial equation becomes
linear in the third-order derivatives; differentiating once with respect to 
the third-order derivatives one gets:

\be
\sum_{(\zeta_{1},\zeta_{2},\zeta_{3})}
\left[ \left(\partial_{D}^{\zeta_{1}\zeta_{2}} \partial_{A}^{\zeta_{3}} -
\partial_{A}^{\zeta_{1}\zeta_{2}} \partial_{D}^{\zeta_{3}} \right) +
2 \left(\partial_{\mu} + \psi^{B}_{\mu} \partial_{B} + 
\psi^{B}_{\mu\rho} \partial^{\rho}_{B}\right) 
\partial^{\zeta_{1}\zeta_{2}}_{D} \partial^{\zeta_{3}\mu}_{A} \right]
{\cal L} = 0.
\label{II3}
\ee

From the initial equation what is left is:

\be
\begin{array}{c}
\partial_{A} {\cal L} - 
(\partial_{\mu} + \psi^{B}_{\mu} \partial_{B}+ 
\psi^{B}_{\mu\rho} \partial_{B}^{\rho})
\partial_{A}^{\mu} {\cal L} + \nonumber \\
(\partial_{\mu} + \psi^{B}_{\mu} \partial_{B}+ 
\psi^{B}_{\mu\rho} \partial_{B}^{\rho}) 
(\partial_{\nu} + \psi^{C}_{\nu} \partial_{C}+ 
\psi^{C}_{\mu\sigma} \partial_{C}^{\sigma}) {\cal L} \equiv 0.
\end{array}
\label{II4}
\ee

So, equation (\ref{trEL2}) is equivalent to the identities 
(\ref{II1})-(\ref{II4}). 

Our strategy in the following will be to find the most general solution of the
equations involving only the second-order derivatives i.e. (\ref{II1}) and
(\ref{II2}). Some comments related to the dependence on the first-order
derivatives will be made at the end. Inspecting more carefully the 
equation (\ref{II2}) it becomes clear that it for 
$
N = 1, 2
$
it follows from (\ref{II1}). Also, (\ref{II1}) for
$
A = B
$
coincides with the same equation for
$
N = 1.
$
This is the reason for studying separately the cases
$
N = 1, 2
$
and
$
N \geq 3.
$

\section{Trivial Second-Order Lagrangians in the Case $N = 1$}

3.1 In this case we will omit completely the field indices 
$
A, B,...
$
because they take only one value. The dependence on the second-order
derivatives is encoded in the equation (\ref{II1}) which becomes in this case: 

\be
\sum_{(\rho_{1},\rho_{2},\rho_{3})}
\partial^{\mu\rho_{1}} \partial^{\rho_{2}\rho_{3}} {\cal L} = 0.
\label{N=1}
\ee

As we have said in the Introduction, we intend to find the most general
solution of this equation using induction over
$n$.
To be able to formulate the induction hypothesis we introduce the following
polynomial expressions, called {\it hyper-Jacobians} \cite{BCO}, \cite{O}
(see also \cite{G1}, \cite{G2}) which in this case have the following form:

\be
J^{\rho_{r+1},...,\rho_{n}}_{\sigma_{1},...,\sigma_{r}} \equiv
\varepsilon^{\rho_{1},...,\rho_{n}} 
\prod_{i=1}^{r} \psi_{\rho_{i}\sigma_{i}} \qquad (r = 0,...,n)
\label{hyperJ}
\ee

We will use consistently Bourbaki conventions:
$
\sum_{\emptyset} \cdots = 0
$
and
$
\prod_{\emptyset} \cdots = 1.
$

We note the following symmetry properties:

\be
J^{\rho_{Q(r+1)},...,\rho_{Q(n)}}_{\sigma_{P(1)},...,\sigma_{P(r)}} = 
(-1)^{|P|+|Q|}
J^{\rho_{r+1},...,\rho_{n}}_{\sigma_{1},...,\sigma_{r}}  
\quad (r = 0,...,n)
\label{antisym}
\ee
where
$P$ 
is a permutation of the numbers
$1,...,r$
and 
$Q$
is a permutations of the numbers 
$r+1,...,n$.

We also note that the following identities are true (see \cite{BCO}):

\be
J^{\rho_{r+1},...,\rho_{n-1},\zeta}_{\sigma_{1},...,\sigma_{r-1},\zeta} = 0
\quad (r = 1,...,n-1).
\label{trace}
\ee

In other words, the hyper-Jacobians are completely antisymmetric in the upper
indices, in the lower indices and are also traceless.

In the following we will need the expression for the derivatives
of the hyper-Jacobians. On easily finds out the following formula:
true:

\be
\begin{array}{c}
\partial^{\mu\nu}
J^{\rho_{r+1},...,\rho_{n}}_{\sigma_{1},...,\nu_{r}}
= \nonumber \\ {1 \over 2} 
\quad \sum_{i=1}^{r} (-1)^{n-i} \delta_{\sigma_{i}}^{\nu} 
J^{\rho_{1},...,\rho_{n},\mu}_{\sigma_{1},...,\hat{\sigma_{i}},...,\sigma_{r}}
+ (\mu \leftrightarrow \nu)
\quad (r = 0,...,n).
\end{array}
\label{derJ}
\ee

This formula suggests the use of the Fock space techniques. Let us emphasize
this point in detail. We will consider the functions
$
J^{\rho_{r+1},...,\rho_{n}}_{\sigma_{1},...,\sigma_{r}}
$
as the components of a tensor
$
\{J_{r}\} \in {\cal H} \equiv {\cal F}^{-}(\R^{n}) \otimes {\cal F}^{-}(\R^{n})
$
where
$
J_{r}
$
belongs to the subspace of homogeneous tensors
$
{\cal H}_{n-r,r}
$
(where
$
{\cal H}_{p,q}
$
is the subspace of homogeneous tensors of degree
$
p, q
$
respectively.) 

We will denote by
$
b^{*(\mu)}, c^{*}_{(\mu)}, b_{(\mu)}, c^{(\mu)}
$
the fermionic creation and respectively the annihilation operators acting in
$
{\cal H}.
$

With these notations one can rewrite (\ref{derJ}) in a more compact way,
namely:

\be
\partial^{\mu\nu} J_{r} = \alpha_{r}
[b^{*(\mu)} c^{(\nu)} + b^{*(\nu)} c^{(\mu)}] 
J_{r-1}; \qquad \alpha_{r} \equiv (-1)^{r-1} {1 \over 2}
\times \sqrt{r \over n-r+1} \qquad (r = 0,...,n).
\label{derJF}
\ee

Also, the identities (\ref{trace}) can be compactly written as follows:

\be
C J_{r} = 0 \qquad (r = 0,...,n)
\label{traceF}
\ee
where we have defined

\be
C \equiv b^{(\mu)} c_{(\mu)}.
\label{constraint}
\ee

We need one more notation for our Fock space machinery, namely
$
<\cdot,\cdot>
$
which is the duality form between
$
{\cal H}
$
and
$
{\cal H}^{*}.
$

3.2 We prove now the main result.

\begin{thm}
The general solution of the equations (\ref{N=1}) is of the following form:

\be
{\cal L} = \sum_{r=0}^{n} 
{\cal L}^{\sigma_{1},...,\sigma_{r}}_{\rho_{r+1},...,\rho_{n}}
J^{\rho_{r+1},...,\rho_{n}}_{\sigma_{1},...,\sigma_{r}}
\label{polyn}
\ee
where the functions
$
{\cal L}^{...}_{...}
$
are independent of
$
\psi_{\mu\nu}
$:

\be
\partial^{\mu\nu}
{\cal L}^{\sigma_{1},...,\sigma_{r}}_{\rho_{r+1},...,\rho_{n}} = 0
\quad (r = 0,...,n),
\ee
and have analogous properties as the hyper-Jacobians, namely the
(anti)symmetry property:

\be
{\cal L}^{\sigma_{P(1)},...,\sigma_{P(r)}}_{\rho_{Q(r+1)},...,\rho_{Q(n)}}
= (-1)^{|P|+|Q|}
{\cal L}^{\sigma_{1},...,\sigma_{r}}_{\rho_{r+1},...,\rho_{n}}
\quad (r = 0,...,n)
\label{antisym-l}
\ee
(where
$P$ 
is a permutation of the numbers
$1,...,r$
and 
$Q$
is a permutations of the numbers 
$r+1,...,n$) 
and also verify the identities:

\be
{\cal L}_{\rho_{r+1},...,\rho_{n-1},\zeta}^{\sigma_{1},...,\sigma_{r-1},\zeta}
= 0
\quad (r = 1,...,n-1)
\label{trace-l}
\ee
(i. e. are traceless). The function coefficients
$
{\cal L}^{...}_{...}
$
are uniquely determined by
$
{\cal L}
$
and the properties (\ref{antisym-l}) and (\ref{trace-l}) above.
\label{structure}
\end{thm}

{\bf Proof:}

It is a particular case of the main theorem from \cite{G2}. It is convenient 
to consider that
$
{\cal L}^{\sigma_{1},...,\sigma_{r}}_{\rho_{r+1},...,\rho_{n}}
$
are the components of a tensor
$
\{{\cal L}^{r}\}
$
in the dual space
$
{\cal H}^{*};
$
explicitly:
$
{\cal L}^{r} \in {\cal H}^{*}_{n-r+1,r}
$
(where
$
{\cal H}^{*}_{p,q}
$
is the subspace of homogeneous tensors of degree
$
p, q
$
respectively.) 

With this trick, formula (\ref{polyn}) can be written in
compact notations as:

\be
{\cal L} = \sum_{r=0}^{n} <{\cal L}^{r},J_{r}>.
\label{compact}
\ee

We will denote by
$
b_{*(\mu)}, c^{*(\mu)}, b^{(\mu)}, c_{(\mu)}
$
the fermionic creation and respectively the annihilation operators acting in
$
{\cal H}^{*}.
$

(i) We now prove the uniqueness statement. So we must show that if

\be
\sum_{r=0}^{n} <{\cal L}^{r},J_{r}> = 0
\label{uniqueness}
\ee
then
$
{\cal L}^{r} = 0 \quad r = 0,...,n.
$

To prove this, we apply to the equation (\ref{uniqueness}) the operator
$
\prod_{i=1}^{p} \partial^{\rho_{i}\sigma_{i}} \quad (p \leq n)
$
and then we will make
$
\psi_{\mu\nu} \rightarrow 0.
$
Using (\ref{derJF}) one easily discovers the following equations:

\be
\prod_{i=1}^{p} \quad \left[ b^{(\rho_{i})} c^{*(\sigma_{i})} +
b^{(\sigma_{i})} c^{*(\rho_{i})} \right] \quad {\cal L}^{p} = 0,
\qquad (p = 0,...,n).
\label{unicity}
\ee

To analyze this system we first define the operator:

\be
{\cal C} \equiv b^{(\rho)} c_{(\rho)}
\ee
and prove by elementary computations that the condition
(\ref{trace-l}) can be rewritten as:

\be
{\cal C} {\cal L}^{r} = 0 \quad (r = 0,...,n).
\label{trace-l-compact}
\ee

At this point it is convenient to define the dual tensors

\be
\tilde{\cal L}^{\sigma_{1},...,\sigma{r};\rho_{1},...,\rho_{r}} \equiv
{(-1)^{r} \over \sqrt{r! (n-r)!}} \varepsilon^{\rho_{1},....\rho_{n}}
{\cal L}^{\sigma_{1},...,\sigma{r}}_{\rho_{r+1},...,\rho_{n}}.
\label{dual}
\ee

Because we have

\be
\tilde{\tilde{\cal L}} = (-1)^{n} {\cal L}
\ee
it is equivalent and more convenient to work in the dual space
$
\tilde{\cal H}.
$
We will denote by
$
b^{*}_{(\mu)}, c^{*}_{(\mu)} 
$
and
$
b^{(\mu)}, c^{(\mu)}
$
the fermionic creation and respectively the annihilation operators acting in
$
\tilde{\cal H}.
$

Then the condition (\ref{trace-l-compact}) rewrites as:

\be
\tilde{\cal C} \tilde{\cal L}^{r} = 0 \qquad (r = 0,...,n)
\label{trace-tilde}
\ee
where 

\be
\tilde{\cal C} \equiv b^{(\mu)} c^{*}_{(\mu)}
\label{constraint-tilde}
\ee
and the equation (\ref{unicity}) becomes:

\be
\prod_{i=1}^{p} \quad \left[ b^{(\rho_{i})} c^{(\sigma_{i})} +
b^{(\sigma_{i})} c^{(\rho_{i})} \right] \tilde{\cal L}^{p} = 0,
\qquad (p = 0,...,n).
\label{unicity-tilde}
\ee

Finally, we need the following number operators:

\be
N_{b} \equiv b^{*(\rho)} b_{(\rho)}; \quad 
N_{c} \equiv c^{*(\rho)} c_{(\rho)}.
\ee

Then one knows that:

\be
N_{b} \vert_{\tilde{\cal H}_{p,q}} = p {\bf 1}, \quad
N_{c} \vert_{\tilde{\cal H}_{p,q}} = q {\bf 1}.
\label{number}
\ee

We analyze the system (\ref{unicity-tilde}) using some simple lemmas.
The proofs are elementary and are omitted.

\begin{lemma}

The following formula is true:

\be
b^{*}_{(\mu)} c^{*}_{(\nu)} \left[ b^{(\mu)} c^{(\nu)} +
b^{(\nu)} c^{(\mu)} \right] = N_{b}  (N_{c} + {\bf 1}) - 
\tilde{\cal C}^{*} \tilde{\cal C}.
\ee
\label{inverse}
\end{lemma}

\begin{lemma}

The operator
$\tilde{\cal C}$
commutes with all the operators of the form

$$
b^{(\mu)} c^{(\nu)} + b^{(\nu)} c^{(\mu)}.
$$ 

Explicitly:

\be
\left[ \tilde{\cal C}, b^{(\mu)} c^{(\nu)} + b^{(\nu)} c^{(\mu)} \right] = 0. 
\ee
\label{commute}
\end{lemma}

\begin{lemma}

If the tensor
$t$
verifies the identity
$
\tilde{\cal C} t = 0
$
the the tensors
$$
\left[ b^{(\mu)} c^{(\nu)} + b^{(\nu)} c^{(\mu)}
\right] t
$$
also verify this identity.
\label{iteration}
\end{lemma}

We now have:

\begin{prop}

Suppose the tensor
$
t \in \tilde{\cal H}_{r,r} \quad (r = 0,...,n)
$
verify the system:

\be
\prod_{i=1}^{p} \quad \left[ b^{(\rho_{i})} c^{(\sigma_{i})} +
b^{(\sigma_{i})} c^{(\rho_{i})} \right] \quad t = 0.
\ee

Then we have
$ 
t = 0.
$
\end{prop}

{\bf Proof:} We apply to this system the operator
$
\prod_{i=1}^{p} b^{*}_{(\rho_{i})} c^{*}_{(\rho_{i})}
$
and make repeated use of the lemmas above.
$\nabla$

The argument involved in the proof above will be called {\it the
unicity argument}.

In conclusion the system (\ref{unicity-tilde}) has the solution
$
\tilde{\cal L}^{p} = 0 \quad (p = 0,...,n).
$

(ii) We start to prove the formula (\ref{polyn}) by induction over
$n$. 
For
$
n = 1
$
the derivation of (\ref{polyn}) is elementary. 
We suppose that we have the assertion of the theorem for a given $n$
and we prove it for
$
n + 1.
$
In this case the indices
$
\mu,\nu, ...
$
takes values (for notational convenience)
$
\mu,\nu, ...= 0,...,n
$
and
$
i,j,...= 1,...,n.
$

If we consider in (\ref{N=1}) that
$
\mu,\rho_{1},\rho_{2},\rho_{3} = 1,...,n
$
then we can apply the induction hypothesis and we get:

\be
{\cal L} = \sum_{r=0}^{n} l^{i_{1},...,i_{r}}_{j_{r+1},...,j_{n}}
I_{i_{1},...,i_{r}}^{j_{r+1},...,j_{n}}.
\label{polyn'}
\ee

Here
$
l^{...}_{...}
$
have properties of the type (\ref{antisym-l}) and ({\ref{trace-l}) and can 
depend on
$
x, \psi^{A}, \psi^{A}_{\mu}
$
{\it and}
$
\psi^{A}_{0\mu}.
$
The expressions
$
I^{...}_{...}
$
are constructed from
$
\psi_{ij}
$
according to the prescription (\ref{hyperJ}).

(iii) We still have at our disposal the relations (\ref{N=1}) where at 
least one index takes the value $0$. The computations are rather easy 
to do using instead of (\ref{polyn'}) the compact tensor notation 
(see (\ref{compact})) and the unicity argument. We give the results directly
for the dual tensors
$
\tilde{l}^{r}.
$

\be
(\partial^{00})^{2} \tilde{l}^{r} = 0 \qquad (r = 0,...,n),
\label{eq1}
\ee

\be
\partial^{00} \partial^{0i} \tilde{l}^{r} = 0 \qquad (r = 0,...,n),
\label{eq2}
\ee

\be
\alpha_{r+1} \left[ b^{(i)} c^{(j)} + b^{(j)} c^{(i)} \right]
\partial^{00} \tilde{l}^{r+1} +
2 \partial^{0i} \partial^{0j} \tilde{l}^{r} = 0 \quad (r = 0,...,n-1),
\label{eq3}
\ee

\be
\partial^{0i} \partial^{0j} \tilde{l}^{n} = 0,
\label{eq4}
\ee

\be
\sum_{(i,j,k)} \left[ b^{(i)} c^{(j)} + b^{(j)} c^{(i)} \right]
\partial^{0k} \tilde{l}^{r} = 0 \qquad (r = 1,...,n).
\label{eq5}
\ee

The expressions
$
\tilde{l}^{r}
$
are obviously considered as tensors from
$
\tilde{\cal H}_{r,r}
$
verifying the restriction:

\be
\tilde{\cal C} \tilde{l}^{r} = 0 \quad (r = 0,...,n).
\label{id}
\ee

As in \cite{G2}, these equations  can be solved i.e. one can describe the
most general solution.

From (\ref{eq1}) we have:

\be
\tilde{l}^{r} = \tilde{l}^{r}_{(0)} + \psi_{00} \tilde{l}^{r}_{(1)}
\label{sol}
\ee
where the functions
$
\tilde{l}^{r}_{(\alpha)} \quad (\alpha = 0,1; \quad r = 0,...,n)
$
verify:

\be
\partial^{00} \tilde{l}^{r}_{(\alpha)} = 0 \quad 
(\alpha = 0,1; \quad r = 0,...,n)
\ee
and also verify identities of the type (\ref{id}):

\be
\tilde{\cal C} \tilde{l}^{r}_{(\alpha)} = 0 \quad (\alpha = 0,1;
\quad r = 0,...,n).
\label{id-alpha}
\ee

From (\ref{eq2})  we also get:
\be
\partial^{0i} \tilde{l}^{r}_{(1)} = 0, \quad (r = 0,...,n)
\label{restr1}
\ee

and finally (\ref{eq3}) - (\ref{eq5}) become:

\be
\alpha_{r+1} \left[ b^{(i)} c^{(j)} + b^{(j)} c^{(i)} \right]
\tilde{l}^{r+1}_{(1)} + 2 \partial^{0i} \partial^{0j} \tilde{l}^{r}_{(0)}
= 0, \qquad (r = 0,...,n-1)
\label{eq3'}
\ee

\be
\partial^{0i} \partial^{0j} \tilde{l}^{n}_{(0)} = 0
\label{4'}
\ee

\be
\sum_{(i,j,k)} \left[ b^{(i)} c^{(j)} + b^{(j)} c^{(i)} \right]
\partial^{0k} \tilde{l}^{r}_{(0)} = 0 \qquad
(r = 0,...,n).
\label{eq5'}
\ee

(iv) We proceed further by applying the operator
$
\partial^{0k}
$
to (\ref{eq3'}); taking into account (\ref{restr1}) we obtain:

\be
\partial^{0i} \partial^{0j} \partial^{0k} \tilde{l}^{r}_{(0)} = 0 
\qquad (r = 0,...,n-1).
\label{eq3''}
\ee

From this relation one obtains a polynomial structure in
$
\psi_{0i}
$
for
$
\tilde{l}^{r}_{(0)} \quad (r = 0,...,n-1):
$

\be
\tilde{l}^{r}_{(0)} = \tilde{l}^{r}_{(00)} + 
\psi_{0i} \tilde{l}^{r}_{(0i)} + {1 \over 2} 
\psi_{0i} \psi_{0j} \tilde{l}^{r}_{(0ij)} \quad (r = 0,...,n-1).
\label{sol1}
\ee

From (\ref{eq4}) one a obtains a similar polynomial structure:
 
\be
\tilde{l}^{n}_{(0)} = \tilde{l}^{n}_{(00)} + 
\psi_{0i} \tilde{l}^{n}_{(0i)} .
\label{sol2}
\ee

Moreover we have the following restrictions on the various tensors appearing in
the preceding two formulae:

\be
\partial^{0i} \tilde{l}^{r}_{(0\mu)} = 0 \quad (r = 0,...,n); \qquad
\partial^{0k} \tilde{l}^{r}_{(0ij)} = 0 \quad (r = 0,...,n-1)
\label{restr1'}
\ee

and

\be
\tilde{\cal C} \tilde{l}^{r}_{(0\mu)} = 0 \quad (r = 0,...,n); \qquad
\tilde{\cal C} \tilde{l}^{r}_{(0ij)} = 0 \quad (r = 0,...,n-1)
\label{id''}
\ee
 
and we also can impose

\be
\tilde{l}^{r}_{(0ij)} = \tilde{l}^{r}_{(0ji)} \quad (r = 0,...,n-1).
\label{restr2}
\ee

If we substitute now (\ref{sol1}) into the original equation (\ref{eq3'})
we obtain

\be
\tilde{l}^{r}_{(0ij)} = - 2 \alpha_{r+1} 
\left[ b^{(i)} c^{(j)} + b^{(j)} c^{(i)} \right]
\tilde{l}^{r+1}_{(1)} \qquad (r = 0,...,n-1).
\label{sol3}
\ee

Finally we substitute the expressions (\ref{sol1}) and (\ref{sol2}) into the
equation (\ref{eq5'}) and we obtain:

\be
\sum_{(i,j,k)} \left[ b^{(i)} c^{(j)} + b^{(j)} c^{(i)} \right]
\tilde{l}^{r}_{(0k)} = 0 \qquad (r = 0,...,n)
\label{eq6}
\ee
and

\be
\sum_{(i,j,k)} \left[ b^{(i)} c^{(j)} + b^{(j)} c^{(i)} \right]
\tilde{l}^{r}_{(0kl)} = 0 \qquad (r = 0,...,n).
\label{eq7}
\ee

One must check that the expression (\ref{sol3}) for 
$
\tilde{l}^{r}_{(0kl)}
$
is compatible with the restrictions (\ref{id''}) by applying by applying 
the operator
$
\tilde{\cal C}
$
to this relation. Also one notes that (\ref{sol3}) identically verifies the
equation (\ref{eq7}).

In conclusion we are left to solve only
(\ref{eq6}) together with the restrictions (\ref{restr1'}),
and (\ref{id''}). We have the following results:

\begin{lemma}

The following formula is valid

\be
\left[ \tilde{\cal C}^{*}, \tilde{\cal C} \right] = N_{b} - N_{c}.
\ee
\end{lemma}

\begin{lemma}

If
$
t \in \tilde{\cal H}_{p,p} 
$
verifies
$
{\cal C} t = 0
$
then it also verifies
$
{\cal C}^{*} t = 0
$
and conversely.
\label{c-star}
\end{lemma}

The proofs of these lemmas are elementary and are omitted. Based on them 
we have

\begin{lemma}

Let
$
t^{k} \in \tilde{\cal H}_{p,p} \quad (k = 1,...,n)
$
be tensors verifying the restriction

\be
\tilde{\cal C} t^{k} = 0 
\label{Ct}
\ee
and the system:

\be
\sum_{(i,j,k)} \left[ b^{(i)} c^{(j)} + b^{(j)} c^{(i)} \right] t^{k} = 0.
\label{permutation}
\ee

Then one can write {\it uniquely}
$t$
in of the following form:

\be
t^{k} = b^{(k)} U + c^{(k)} V
\label{T}
\ee
with
$
U \in \tilde{\cal H}_{p+1,p}
$
and
$
V \in \tilde{\cal H}_{p,p+1}
$
verifying

\be
\tilde{\cal C} U = V \quad \tilde{\cal C}^{*} U = 0 \quad 
\tilde{\cal C} V = 0 \quad \tilde{\cal C}^{*} V = U.
\label{UV}
\ee

Here we put by convention
$
\tilde{\cal H}_{p,q} \equiv \{0\}
$
if at least one of the indices
$p$
and
$q$
is negative
or
$n+1$.
\label{permutation-lemma}
\end{lemma}

{\bf Proof:}
We apply to the equation (\ref{permutation}) the operator
$
b^{*}_{(i)} c^{*}_{(j)}
$
and we find out (after summation over $i$ and $j$ and taking
into account (\ref{Ct}):

\be
(p+2)t^{k} = b^{(k)} b^{*}_{(l)} t^{l} + c^{(k)} c^{*}_{(l)} t^{l}.
\ee

So we have the formula from the statement with:
$
U = (p+2)^{-1} b^{*}_{(l)} t^{l}
$
and
$
V = (p+2)^{-1} c^{*}_{(l)} t^{l}.
$

These expressions verify the identities (\ref{UV}). Conversely, if 
we have (\ref{T}) and (\ref{UV}) it remains to check that the equations 
(\ref{permutation}) and (\ref{Ct}) are indeed identically satisfied.
$\nabla$

From this lemma one can write  down that the most general solution of
(\ref{eq6}). Combining with the previous results one obtains the most general
expression for the tensors
$
\tilde{l}^{r}.
$

Reverting to the original tensors
$
l^{r}
$
one obtains easily that the most general expression for them is:

\be
l^{r} = l^{r}_{(0)} + \psi_{0i} \left[ b^{(i)} U^{r} + c^{*(i)} V^{r}
\right] - 2 \alpha_{r+1} \psi_{0i} \psi_{0j} b^{(i)} c^{*(j)}
l^{r+1}_{(1)} + \psi_{00} l^{r}_{(1)} \qquad (r = 0,...,n).
\label{sol-gen}
\ee

The tensors 
$
l^{r}_{(0)}, l^{r}_{(1)} \in \tilde{\cal H}_{r,n-r},
U^{r} \in \tilde{\cal H}_{r+1,n-r}, V^{r} \in \tilde{\cal H}_{r,n-r-1}
$
are not completely arbitrary; they must satisfy the following relations:

\be
{\cal C} l^{r}_{(\alpha)} = 0, \quad (\alpha = 0,1; \quad r = 0,...,n),
\label{iden1}
\ee

\be
{\cal C} U^{r} = V^{r}, \quad {\cal C} V^{r} = 0, \quad
{\cal C}^{*} U^{r} = 0, \quad {\cal C}^{*} V^{r} = U^{r} \quad
(r = 0,...,n)
\label{iden2}
\ee

and

\be
\partial^{\mu\nu} l^{r}_{(\alpha)} = 0, \quad
\partial^{\mu\nu} U^{r} = 0, \quad 
\partial^{\mu\nu} V^{r} = 0, \qquad
(r = 0,...,n; \quad \alpha = 0,1).
\label{iden3}
\ee

The structure of the tensors
$
l^{r} \quad (r = 0,...,n)
$
is completely elucidated.

(v) It remains to introduce these expressions for
$
l^{r}
$
in (\ref{polyn'}) and regroup the terms. Like in \cite{G1}, \cite{G2}
one obtains the desired formula (\ref{polyn}) for
$
n+1
$
with the tensors
$
{\cal L}^{r}
$
expressed in terms of the tensors defined in the proof above. Finally one 
must check that the tensors 
$
{\cal L}^{r}
$
also verify the induction hypothesis i.e. the identities (\ref{trace-l}). 
This is done after some computations using (\ref{iden1}) - (\ref{iden3}) 
and the induction is finished.
\qed

\begin{rem}

We make a last comment concerning the unicity statement from the proof. First,
the non-uniqueness is easy to explain because if one add to the tensors
$
{\cal L}^{...}_{...}
$
contributions containing at least a factor
$
\delta_{\rho_{j}}^{\sigma_{i}}
$
then it immediately follows from the identity (\ref{trace}) that the right hand
side of the formula (\ref{polyn}) is not changed. So, the constrain
(\ref{trace-l}) is a way of eliminating these type of contributions respecting
in the same time the antisymmetry properties of the functions
$
{\cal L}^{...}_{...}
$ 
i.e. to obtain the {\it traceless} part of
$
{\cal L}^{...}_{...}.
$

In this context we mention that such a decomposition of  a tensor in a
traceless part and a rest containing at least a delta factor is true in
extremely general conditions as it is proved in \cite{K3}.
\end{rem}

3.2 Let us prepare the ground for the analysis of the more complicated case
$
N \geq 2.
$
First we note that if in analogy to (\ref{dual}) we define:

\be
\tilde{J}_{\sigma_{1},...,\sigma_{r};\rho_{1},...,\rho_{r}} \equiv
{(-1)^{r} \over \sqrt{r! (n-r)!}} \varepsilon_{\rho_{1},....\rho_{n}}
J_{\sigma_{1},...,\sigma_{r}}^{\rho_{r+1},...,\rho_{n}}.
\label{dual-hyperJ}
\ee
then can rewrite (\ref{polyn}) as follows:

\be 
{\cal L} = \sum_{r=0}^{n}
\tilde{\cal L}^{\sigma_{1},...,\sigma_{r};\rho_{1},...,\rho_{r}}
\tilde{J}_{\sigma_{1},...,\sigma_{r},\rho_{1},...,\rho_{r}}.
\label{polyn-tilde}
\ee

We intend to use equation (\ref{II1}) first for the case
$
A = B = 1,2...,N.
$
It is clear that we will be able to apply the theorem above. To do this we
define in analogy to (\ref{hyperJ}) and (\ref{dual-hyperJ}) the expressions

\be 
J^{(A)\rho_{r+1},...,\rho_{n}}_{\sigma_{1},...,\sigma_{r}} \equiv
\varepsilon^{\rho_{1},...,\rho_{n}} 
\prod_{i=1}^{r} \psi_{\rho_{i}\sigma_{i}}^{A} \qquad (r = 0,...,n; 
\quad A = 1,...,N)
\label{hyperJN}
\ee
and

\be
\tilde{J}^{(A)}_{\sigma_{1},...,\sigma_{r};\rho_{1},...,\rho_{r}} \equiv
{(-1)^{r} \over \sqrt{r! (n-r)!}} \varepsilon_{\rho_{1},....\rho_{n}}
J_{\sigma_{1},...,\sigma_{r}}^{(A)\rho_{r+1},...,\rho_{n}}.
\label{dual-hyperJN}
\ee

Then the equations (\ref{II1}) for
$
A = B
$
will produce an expression of the following form:

\be
{\cal L} = \sum_{r,s,...,=0}^{n} 
{\cal L}^{\sigma_{1},...,\sigma_{r};\mu_{1},...,\mu_{s};
\cdots}_{\rho_{r+1},...,\rho_{n};\nu_{s+1},...,\nu_{n};\cdots}
J^{\rho_{r+1},...,\rho_{n}}_{\sigma_{1},...,\sigma_{r}}
J^{\nu_{s+1},...,\nu_{n}}_{\mu_{1},...,\mu_{s}} \cdots
\label{polynN}
\ee
where the functions
$
{\cal L}^{...}_{...}
$
are verifying the following properties:

\be
\partial^{\mu\nu}_{A} {\cal L}^{...}_{...} = 0
\ee

\be
{\cal L}^{\cdots;\sigma_{P(1)},...,\sigma_{P(r)};\cdots}_{\cdots;
\rho_{Q(r+1)},...,\rho_{Q(n)};\cdots} = 
(-1)^{|P|+|Q|} 
{\cal L}^{\cdots;\sigma_{1},...,\sigma_{r};\cdots}_{\cdots;
\rho_{r+1},...,\rho_{n};\cdots} 
\ee
(where
$P$
is a permutation of the numbers 
$
1,...,r
$
and
$Q$
is a permutation of the numbers
$
r+1,...,n
$)
and

\be
{\cal L}^{\cdots;\sigma_{1},...,\sigma_{r-1},\zeta;\cdots}_{\cdots;
\rho_{r+1},...,\rho_{n-1},\zeta;\cdots} = 0.
\ee

Again the analysis is much simplified if one uses tensor notations.
Generalizing  in an obvious way the scalar case the functions
$
{\cal L}^{...}_{...}
$
will become the components of a tensor
$
{\cal L} \in ({\cal H}^{*})^{\otimes N}
$
and one can write (\ref{polynN}) in a more compact manner:

\be
{\cal L} = \sum_{r_{1},...,r_{N}=1}^{n} 
<{\cal L}^{r_{1},...,r_{N}}, J^{(1)}_{r_{1}} \otimes \cdots J^{(N)}_{r_{N}}> 
\label{polynN-compact}
\ee 
where 
$
J^{(1)}_{r_{1}} \otimes \cdots J^{(N)}_{r_{N}} \in 
{\cal H}^{\otimes N}
$
and
$
<\cdot,\cdot>
$
is the duality form.

Let 
$
b^{*(\mu)}_{(A)}, c^{*(A)}_{(\mu)}, b^{(A)}_{(\mu)},c_{(A)}^{(\mu)}
$
be the creation and the annihilation operators acting in
$
{\cal H}^{\otimes N}
$
and
$
b^{*(A)}_{(\mu)}, c^{*(\mu)}_{(A)}, b^{(\mu)}_{(A)},c_{(\mu)}^{(A)}
$
the corresponding operators from
$
({\cal H}^{*})^{\otimes N}.
$

Then the constraints of the type (\ref{trace-l-compact}) can be written as
follows: 

\be
{\cal C}_{A} {\cal L}^{r_{1},...,r_{N}} = 0 \quad (A = 1,...,N)
\label{trace-lN}
\ee
where we have defined:

\be
{\cal C}_{A} \equiv b^{(\mu)}_{(A)} c_{(\mu)}^{(A)} \quad (A = 1,...,N).
\ee

The expressions of the type (\ref{polynN}) or (\ref{polynN-compact}) are unique
in the sense that
$
{\cal L}
$
uniquely determines the function coefficients
$
{\cal L}^{r_{1},...,r_{N}};
$
this follows directly from the uniqueness statement of theorem \ref{structure}.

It is convenient to work with the dual tensors
$
\tilde{\cal L}^{r_{1},...,r_{N}} \in \tilde{\cal H}^{\otimes N} 
\quad (A = 1,...,N)
$
defined analogously as in (\ref{dual}) which will verify the constraints:
\be
\tilde{\cal C}_{A} \tilde{\cal L}^{r_{1},...,r_{N}} = 0 \quad (A = 1,...,N)
\label{C}
\ee
where
\be
\tilde{\cal C}_{A} \equiv b^{(\mu)}_{(A)} c_{(\mu)}^{*(A)} \quad (A = 1,...,N)
\ee
are the expressions of the constraints in the dual space.

Our goal in the next two sections will be to prove the following result:

\begin{thm}

The most general solution of the equations (\ref{II1}) and (\ref{II2}) is of
the form:

\be
{\cal L} = \sum_{r_{1},...,r_{N}=1}^{n} 
<\tilde{\cal L}^{r_{1},...,r_{N}}, 
\tilde{J}^{(1)}_{r_{1}} \otimes \cdots \tilde{J}^{(N)}_{r_{N}}>.
\label{polynN-compact-dual}
\ee 

The tensors
$
\tilde{\cal L}^{\sigma_{1},...,\sigma_{r};\rho_{1},...,\rho_{r};
\mu_{1},...,\mu_{s};\nu_{1},...,\nu_{s};\cdots}
$
verify, the usual antisymmetry and tracelessness properties, but moreover
they verify the property of complete antisymmetry in {\it all} the indices
$
\sigma_{1},...,\sigma_{r},\mu_{1},...,\mu_{s},...
$
i.e. they verify the identities

\be
\left[ b^{(\mu)}_{(A)} b^{(\nu)}_{(B)} + (\mu \leftrightarrow \nu) \right] 
\tilde{\cal L}^{r_{1},...,r_{N}} = 0.
\label{antisymmetry}
\ee
\label{structureN}
\end{thm}

To do this we will use the remaining equations i.e. (\ref{II1}) for
$
A \not= B
$
and (\ref{II2}). Using the compact expression (\ref{polynN-compact}) one
obtains from (\ref{II1}):

\be
\sum_{(\rho_{1},\rho_{2},\rho_{3})} \left[ 
b^{(\mu)}_{(A)} c^{(\rho_{1})}_{(A)} + b^{(\rho_{1})}_{(A)} c^{(\mu)}_{(A)}
\right] \left[
b^{(\rho_{2})}_{(B)} c^{(\rho_{3})}_{(B)} + 
b^{(\rho_{3})}_{(B)} c^{(\rho_{2})}_{(B)} \right]
\tilde{\cal L}^{r_{1},...,r_{N}} + (A \leftrightarrow B) = 0
\label{II1-Fock}
\ee

and from (\ref{II2}) it follows:

\be
\sum_{(\rho_{1},\rho_{2},\rho_{3})} \sum_{(\zeta_{1},\zeta_{2},\zeta_{3})}
\left[  b^{(\zeta_{1})}_{(D)} c^{(\zeta_{2})}_{(D)} + 
b^{(\zeta_{2})}_{(D)} c^{(\zeta_{1})}_{(D)} \right] 
\left[ b^{(\rho_{1})}_{(B)} c^{(\rho_{2})}_{(B)} + 
b^{(\rho_{2})}_{(B)} c^{(\rho_{1})}_{(B)} \right] 
\left[ b^{(\zeta_{3})}_{(A)} c^{(\rho_{3})}_{(A)} + 
b^{(\rho_{3})}_{(A)} c^{(\zeta_{3})}_{(A)} \right] 
\tilde{\cal L}^{r_{1},...,r_{N}} = 0
\label{II2-Fock}
\ee

For the case 
$
N = 2
$
to be analysed in the next Section it easily follows that (\ref{II2-Fock})
follows from (\ref{II1-Fock}), so we will have to analyse only the first
equation for
$
A \not= B.
$
For the case
$
N \geq 3
$
to be analysed in Section 5 it will be convenient to start with 
(\ref{II2-Fock}).

\section{Trivial Second-Order Lagrangians for the Case $N = 2$}

As we have said before we analyse (\ref{II1-Fock}) in the case
$
A \not= B.
$

It is convenient to redefine 
$
b^{(\mu)}_{(A)} \rightarrow d^{\mu}, c^{(\mu)}_{(A)} \rightarrow e^{\mu},
b^{(\mu)}_{(B)} \rightarrow b^{\mu}, c^{(\mu)}_{(B)} \rightarrow c^{\mu}.
$

Then the equation (\ref{II1-Fock}) takes the generic form:

\be
\sum b^{\mu} c^{\nu} d^{\rho} e^{\sigma} t = 0
\label{II1-compact}
\ee
where the sum runs over all the permutations of indices
$
\mu, \nu, \rho, \sigma
$
and 
$t$
is an arbitrary tensor from
$
\tilde{\cal H}_{k,k',r,r}.
$
Here
$k, k', r, r$ are the eigenvalues of the operators
$
N_{b}, N_{c}, N_{d} 
$
and
$
N_{e}
$
respectively. We now define the following operators:

\be
A \equiv c^{\rho} d^{*}_{\rho}, \quad
B \equiv d^{\rho} e^{*}_{\rho}, \quad
C \equiv e^{\rho} c^{*}_{\rho}, \quad
Z \equiv b^{\rho} c^{*}_{\rho}
\label{ABC}
\ee
and there are some constraints to be taken into account, namely (see (\ref{C})
and lemma \ref{c-star}):

\be
B t = 0, \quad B^{*} t = 0
\label{B}
\ee

and

\be
Z t = 0, \quad Z^{*} t = 0.
\label{Z}
\ee

We will use in the following only the constraint (\ref{B}). We start the proof
of theorem \ref{structureN} by a series of lemmas and propositions.

\begin{lemma}

If the tensor
$t$
verifies the equation (\ref{II1-compact}) and the constraints (\ref{B}) then it
also verifies the equation
\be
(r+2) \left[ (k+1) (r+3) {\bf 1} - M \right] b^{\mu} t =
c^{\mu} U_{(0)} + d^{\mu} V_{(0)} + e^{\mu} W_{(0)}
\label{II1-contraction}
\ee
where

\be
M \equiv A^{*} A + C C^{*}
\label{M}
\ee

and
$
U_{(0)}, V_{(0)}, W_{(0)}
$
are some tensors constructed from
$t$. (We will not need in the following their explicit expressions.)
\end{lemma}

{\bf Proof:} One applies to the equation (\ref{II1-compact}) the operator
$
c^{*}_{\nu} d^{*}_{\rho} e^{*}_{\sigma}
$
and uses the constraints (\ref{B}).
$\nabla$

We want to prove that the operator in the square brackets from
(\ref{II1-contraction}) is invertible. To do this we notice that 
$M$
can be restricted to the subspace of 
$
\tilde{\cal H}_{k,k',r,r}
$
determined by
$
Ker{B} \cap Ker(B^{*}).
$
We denote this subspace by
$h$
and by
$M'$
the restriction of
$M$
to
$h$.
Then we have

\begin{prop}

The spectrum of the operator
$M'$
is described by:

\be
{\it Spec}(M') \subset \{ v(v+r-k+2) | v = 0,...,v_{0}\}
\label{spectrum}
\ee
where
$
v_{0} \equiv min\{k, 2(n-r)\}.
$
In particular we have 

\be
{\it Spec}(M') \subset [0, k (r+2)].
\label{S}
\ee
\label{spectrum-M}
\end{prop}

{\bf Proof:} 

One finds out after some tedious but straightforward computations that if 
the tensor
$t$
verifies the eigenvalue equation

\be
M' t = \lambda t
\label{X}
\ee

then one also has

\be
M' A^{s} (C^{*})^{u} t = \lambda_{s,u} A^{s} (C^{*})^{u} t.
\label{eigen}
\ee

Here 
$s$
and
$u$
are natural numbers verifying
$
s, u \leq n-r, \quad s+u \leq k
$
and the expression for the eigenvalue is

\be
\lambda_{s,u} \equiv \lambda - \Lambda_{s+u}
\ee
where

\be
\Lambda_{v} \equiv v(v-r-k+2).
\ee

The proof of the formula above is most easily done by induction: first one
considers the case
$
s = 0
$
and use the induction over
$u$
and next one proves (\ref{eigen}) by induction over
$s$.
Needless to say, one must make use of the various anticommutation relations
between the operators
$A, B, C$
and their Hermitian conjugates. 

Now one supposes that
$
\lambda \not\in {\it Spec}(M')
$
i.e.
$
\lambda \not= \Lambda_{v} \quad (v= 0,1,...,v_{0}).
$
In this case it follows that
$
\lambda_{s,u} \not= 0
$
and we will be able to prove that one has

\be
A^{s} (C^{*})^{u} t = 0 \quad (s,u \leq n-r, \quad s+u \leq k).
\label{recurrence}
\ee

We analyse separately two cases. If
$
2r + k \leq 2n
$
we can take in (\ref{eigen})
$s$
and
$u$
such that
$
s + u = k.
$

One obtains that

\be
\lambda_{s,u} A^{s} (C^{*})^{u} t = 0.
\ee

Because
$
\lambda_{s,u} \not= 0
$
we have (\ref{recurrence}) for
$
s + u = k.
$

Next, one proves this relation for
$
s + u \leq k
$
by recurrence (downwards) over
$
v = s + u
$
using again (\ref{eigen}). In the case
$
2r + k > 2n
$
the proof of (\ref{recurrence}) is similar, only one has to start the induction
downwards from
$
s + u = 2(n-r).
$

The relation (\ref{recurrence}) is proved. If we take in this relation
$
s = u = 0
$
we get
$
t = 0.
$

In conclusion, if
$
\lambda  
$
does not belong to the set
$
{\it Spec}(M')
$
then the equation (\ref{X}) does not have non-trivial solutions. Because the 
operator
$M'$
lives in a finite dimensional Hilbert space the first assertion of the 
proposition follows. The second assertion follows from the first and from
$
M \geq 0.
$
$\qed$

\begin{cor}

The matrix
$
(k+1) (r+3) {\bf 1} - M'
$
is invertible.
\end{cor}

{\bf Proof:} 
$
(k+1) (r+3)
$
does not belong to the spectrum of 
$M'$.
$\nabla$

Now we come back to the equation (\ref{II1-contraction}); using the corollary
above and the finite dimensional functional calculus, it is not hard to prove
that one obtains from this equation the following consequence:

\be
b^{\mu} t = c^{\mu} U + d^{\mu} V + e^{\mu} W
\label{UVW}
\ee 
where
$
U, V, W
$
are some tensors verifying

\be
B U = 0, \quad B V = W, \quad B W = 0, \quad B^{*} U = 0, \quad B^{*} V = 0, 
\quad B^{*} W = V.
\ee

A structure formula of the type (\ref{UVW}) is valid for every tensor
$
\tilde{\cal L}^{r,k}
$
appearing in the structure formula for the trivial Lagrangian. It it important
to note that in deriving this result we have used only the constraints 
(\ref{B}) and not the constraints (\ref{Z}). So, the tensors
$
\tilde{\cal L}^{r,k}
$
are not uniquely fixed by the unicity argument. We use the possibility of
redefining these tensors to our advantage. Indeed, if one inserts formulas of
the type (\ref{UVW}) into the expression of the Lagrangian one can show that
the contribution following from the first term is null (one must use the
tracelessness of the hyper-Jacobians). In other words, one can redefine the 
tensors
$
\tilde{\cal L}^{r,k}
$
such that one has:

\be
(r+2) b^{\mu} \tilde{\cal L}^{r,k} = d^{\mu} V + e^{\mu} W.
\ee

Now one can make this formula more precise is one uses in a clever way lemma 
\ref{permutation-lemma}, namely the following relation stands true:

\be
(r+2) b^{\mu} \tilde{\cal L}^{r,k} = (d^{\mu} {\cal D} + e^{\nu} {\cal E})
{\cal L}^{r,k}
\label{L}
\ee
where we have introduce new notations:

\be
{\cal D} \equiv b^{\mu} d^{*}_{\mu}, \quad {\cal E} \equiv b^{\mu} e^{*}_{\mu}.
\ee

Now we have

\begin{lemma}

The following formula is valid:

\be
{(r+p+1)! \over (r+1)!} b^{\mu_{1}} \cdots b^{\mu_{p}} \tilde{\cal L}^{r,k} = 
(-1)^{[p/2]} \sum_{s=0}^{p} (-1)^{(p+1)s} C^{s}_{p}
{\cal A}_{p} (d^{\mu_{1}} \cdots d^{\mu_{s}} e^{\mu_{s+1}} \cdots e^{\mu_{p}}) 
{\cal D}^{s} {\cal E}^{p-s} \tilde{\cal L}^{r,k}.
\label{bbb}
\ee

Here
$
[m]
$
is the integer part of
$m$,
$
C^{s}_{p} \equiv {p! \over s! (p-s)!}
$
and
$
{\cal A}_{p}
$
is the operator of antisymmetrization in the indices
$
\mu_{1},...,\mu_{p}.
$
\end{lemma}

{\bf Proof:} By very long computations using induction over
$p$.
Indeed, for 
$
p = 0
$
the formula is trivial and for
$
p = 1
$
we have (\ref{L}). 
$\nabla$
 
In particular, if we take in (\ref{bbb})
$
p = k
$
one obtains after some prelucrations

\be
{(r+k+1)! \over (r+1)!} b^{\mu_{1}} \cdots b^{\mu_{k}} \tilde{\cal L}^{r,k} = 
\sum_{s=0}^{k} (-1)^{(k+1)s+[s/2]} {1 \over (k-s)!}
{\cal A}_{p} (d^{\mu_{1}} \cdots d^{\mu_{s}} e^{\mu_{s+1}} \cdots e^{\mu_{k}}) 
B^{k-s} L^{r,k} 
\label{bcd}
\ee
where
$
L^{r,k} \in \tilde{\cal H}_{r+k,r,0,k}
$
is given by

\be
L^{r,k} = {\cal D}^{k} \tilde{\cal L}^{k,r}.
\ee

Using indices the expression of 
$
{\cal L}
$
becomes

\be
\begin{array}{c}
{\cal L} = \sum_{r,k=0}^{n} {(r+1)! \over (r+k+1)!} \sum_{s=0}^{k} 
(-1)^{(k+1)s+[s/2]} {1 \over (k-s)!} 
(B^{k-s} L)^{\mu_{1},...\mu_{s}\rho_{1},...,\rho_{r};
\mu_{s+1},...,\mu_{k}\sigma_{1},...,\sigma_{r};\emptyset;\nu_{1},...,\nu_{k}}
\nonumber \\
\tilde{J}^{(1)}_{\rho_{1},...,\rho_{r};\sigma_{1},...,\sigma_{r}}
\tilde{J}^{(2)}_{\mu_{1},...,\mu_{k};\nu_{1},...,\nu_{k}}
\end{array}
\ee

Now one uses the explicit expression for the operator
$B$
and the tracelessness of the hyper-Jacobians to show that one can replace in
the formula above 
$
B^{k-s} 
$
by a sum of lower powers of 
$B$. 
In the end one finds out by recurrence that the sum over 
$s$
disappears and the formula above is transformed into:

\be
{\cal L} = \sum_{r,k=0}^{n} {(r+1)! \over (r+k+1)!} (-1)^{k(k-1)/2} 
L^{\mu_{1},...\mu_{k}\rho_{1},...,\rho_{r};
\sigma_{1},...,\sigma_{r};\emptyset;\nu_{1},...,\nu_{k}}
\tilde{J}^{(1)}_{\rho_{1},...,\rho_{r};\sigma_{1},...,\sigma_{r}}
\tilde{J}^{(2)}_{\mu_{1},...,\mu_{k};\nu_{1},...,\nu_{k}}.
\label{structureII}
\ee

In other words, by redefining
$
{\cal L}^{r,k} \rightarrow {\cal L}^{r,k}_{1}
$
where:

\be
{\cal L}_{1}^{\rho_{1},...,\rho_{r};\sigma_{1},...,\sigma_{r};
\mu_{1},...\mu_{k};\nu_{1},...,\nu_{k}} \equiv
(-1)^{k(k-1)/2} {(r+1)! \over (r+k+1)!}
L^{\mu_{1},...\mu_{k}\rho_{1},...,\rho_{r};
\sigma_{1},...,\sigma_{r};\emptyset;\nu_{1},...,\nu_{k}}
\label{redefine}
\ee

one preserves the formula (\ref{structureII}) and has moreover

\be
(b^{\mu} d^{\nu} + b^{\nu} d^{\mu}) {\cal L}^{r,k}_{1} = 0.
\ee
 
This observation finishes the proof of the theorem \ref{structureN} for the 
case
$
N = 2.
$

\section{Trivial Second-Order Lagrangians in the Case $N \geq 3$}

In this case we start with the equation (\ref{II2-Fock}) and  note that it
gives something non-trivial {\it iff} all the three indices
$
A, B, D
$
are distinct one of the other. In this case it is convenient to redefine
$$
b^{(\mu)}_{(D)} \rightarrow d^{\mu},
c^{(\mu)}_{(D)} \rightarrow e^{\mu},
b^{(\mu)}_{(B)} \rightarrow f^{\mu},
c^{(\mu)}_{(B)} \rightarrow g^{\mu},
b^{(\mu)}_{(A)} \rightarrow b^{\mu},
c^{(\mu)}_{(A)} \rightarrow c^{\mu}
$$
and to obtain an equation of the following type:

\be
\sum_{(\rho_{1},\rho_{2},\rho_{3})} \sum_{(\zeta_{1},\zeta_{2},\zeta_{3})}
\left(  d^{\zeta_{1}} e^{\zeta_{2}} + 
d^{\zeta_{2}} e^{\zeta_{1}} \right)
\left( f^{\rho_{1}} g^{\rho_{2}} + 
f^{\rho_{2}} g^{\rho_{1}} \right)
\left( b^{\zeta_{3}} c^{\rho_{3}} + 
b^{\rho_{3}} c^{\zeta_{3}} \right) t = 0
\label{II2-compact}
\ee
where
$
t \in \tilde{\cal H}_{r,r,t,t,k,k};
$
here
$
r_{D} = r, r_{B} = t, r_{A} = k.
$

Moreover, the tensor 
$t$
must satisfy the constraints (\ref{B}) and (\ref{Z}). It is convenient to
define 

\be
t_{1} \equiv \sum_{(\rho_{1},\rho_{2},\rho_{3})} 
\left( f^{\rho_{1}} g^{\rho_{2}} + 
f^{\rho_{2}} g^{\rho_{1}} \right) b^{\rho_{3}} t
\label{tensor-X}
\ee

and

\be
t_{2} \equiv \sum_{(\rho_{1},\rho_{2},\rho_{3})} 
\left( f^{\rho_{1}} g^{\rho_{2}} + 
f^{\rho_{2}} g^{\rho_{1}} \right) c^{\rho_{3}} t.
\label{tensor-Y}
\ee

Let us note that
$
t_{1} \in \tilde{\cal H}_{r,r,t-1,t-1,k-1,k}
$
and
$
t_{2} \in \tilde{\cal H}_{r,r,t-1,t-1,k,k-1}.
$

Then we can rewrite (\ref{II2-compact}) in equivalent way if we use lemma 
\ref{permutation-lemma} and the constraint (\ref{B}), namely:

\be
(r+2) (c^{\mu} t_{1} + b^{\mu} t_{2}) = 
d^{\mu} (A t_{1} + {\cal D} t_{2}) + c^{\mu} (C^{*} t_{1} + {\cal E} t_{2}).
\label{II2-XY}
\ee

We can formulate now the central result of this Section:

\begin{prop}

The equation (\ref{II2-XY}) above together with the constrains (\ref{B}) and
(\ref{Z}) have only the trivial solution
$
t_{1} = 0.
$
\end{prop}

{\bf Proof:} The proof is extremely tedious and we give only a minimal number
of details. 

(i) We apply to the equation (\ref{II2-XY}) the operator
$
c^{*}_{\mu}.
$
Let us define the following operators

\be
Q \equiv A^{*} {\cal D} + C {\cal E}
\label{Q}
\ee

and

\be
R \equiv M + Q Z^{*}.
\label{R}
\ee

Then one can write the result in a very simple form:

\be
R t_{1} = (r+2) (k+1) t_{1} 
\label{RX}
\ee

so one must investigate this eigenvalue problem.

(ii) We denote by
$h'$
the subspace of
$
\tilde{\cal H}_{r,r,t-1,t-1,k-1,k}
$
defined by
$
Ker(B) \cap Ker(B^{*}) \cap Ker(Z)
$
and note that
$
t_{1} \in h'.
$

Next one shows by simple computations that the operator
$R$
commutes with the operators
$
B, B^{*}, Z
$
so it makes sense to restrict it to the subspace
$h'.$
The same assertion is true with respect to the operator
$R^{*}.
$
We will denote these restrictions by
$R'$
and
$R'^{*}$
respectively. From this observations it follows in particular that the operator
$
R - R^{*}
$
leaves the subspace
$h'$
invariant. But one computes that

\be
R - R^{*} = Q Z^{*} - Z Q^{*} = Z^{*} Q - Q^{*} Z
\label{RQZ}
\ee

and from the last equality it easily follows that

\be
(t_{1}, (R - R^{*}) t_{1}') = 0 \quad \forall t_{1}, t_{1}' \in h'.
\ee 

Because
$
R - R^{*}
$ 
leaves the subspace
$h'$
invariant we conclude that

\be
R' - R'^{*} = 0.
\ee

Combining with (\ref{RQZ}) one discovers that:

\be
(Q Z^{*} - Z Q^{*}) t_{1} = 0 \quad \forall t_{1} \in h'.
\ee

Applying to this relation the operator
$Z$
produces after some computations:

\be
Q t_{1} = 0.
\label{QX}
\ee

Let us introduce the following notation:

\be
N \equiv {\cal D}^{*} {\cal D} + {\cal E}^{*} {\cal E}.
\label{N}
\ee

Then, taking into account (\ref{QX}) one can obtain from (\ref{RX}) the
following relation:

\be
(2M - N) t_{1} = (r+2) (k+1) t_{1}.
\label{MN}
\ee

(iii) In the same way one can obtain similar results concerning the tensor
$t_{2}$.
More precisely one denotes by
$h''$
the subspace of
$
\tilde{\cal H}_{r,r,t-1,t-1,k,k-1}
$
defined by
$
Ker(B) \cap Ker(B^{*}) \cap Ker(Z^{*})
$
and notes that
$
t_{2} \in h''.
$
Then one shows that similarly to (\ref{QX}) and (\ref{MN}) one has:

\be
Q^{*} t_{2} = 0
\label{Q-star}
\ee

and

\be
(2N - M) t_{2} = (r+2) (k+1) t_{2}.
\label{NM}
\ee

(iv) The relation (\ref{QX}) suggests to investigate if it is possible to
restrict the operator
$
2M - N
$
to the subspace
$
h_{Q} \equiv h' \cap Ker(Q).
$

One shows by elementary but rather long computations that this assertion is
true for the operators
$M, N$
so it makes sense to define by
$
M_{Q}, N_{Q}
$
the corresponding restrictions. So, indeed the operator
$
2M - N
$
leaves the subspace
$h_{Q}$
invariant. 

(v) Moreover one computes the commutator of
$M$
and
$N$
and shows that:

\be
[M_{Q}, N_{Q}] = 0.
\ee

It follows that the operators
$
M_{Q}, N_{Q}
$
can be simultaneously diagonalized. Let us suppose that the equation (\ref{MN})
can have a non-trivial solution. Then it is easy to show that there must exists
at least one non-trivial solution 
$t_{1}$
of this equation such that
$t_{1}$
is a simultaneous eigenvalue of the operators
$M, N.$
Explicitly:

\be
M t_{1} = \lambda t_{1}, \quad N t_{1} = \mu t_{1}
\label{MNX}
\ee

and

\be
2 \lambda - \mu = (k+1) (r+2).
\ee

(vi) Let us notice now that the relation (\ref{Q-star}) can be written as 
follows:

\be
Q^{*} Z^{*} t_{1} = 0.
\ee

One applies to this relation, first the operator
$Z$
and next the operator
$Q$;
after some computations one gets:

\be
(M - N)^{2} t_{1} = [Q, Q^{*}] t_{1}.
\ee

One evaluates explicitly the commutator and uses (\ref{MNX}) to obtain the
following restriction on the eigenvalue
$
\lambda
$:

\be
(\lambda - \lambda_{0})^{2} + (r-k+2) (\lambda - \lambda_{0}) + \lambda = 0
\label{lambda}
\ee

where we have denoted for simplicity

\be
\lambda_{0} \equiv (k+2) (r+1)
\ee

i.e. the eigenvalue appearing in the right hand side of the equation
(\ref{MN}). Now it is easy to prove  that the solutions of the equation 
(\ref{lambda}) (if they exist at all) must be greater than
$
k (r+2).
$
But this conflicts with Proposition \ref{spectrum-M} (see formula (\ref{S}))
so we conclude that the equation (\ref{MN}) have only the trivial solution in
the subspace
$
h_{Q}.
$
This finishes the proof of the similar assertion from the statement.
$\qed$

Remembering the definition of the tensor
$t_{1}$
we just have find out that we have

\be
\sum_{(\rho_{1},\rho_{2},\rho_{3})} 
\left( f^{\rho_{1}} g^{\rho_{2}} + 
f^{\rho_{2}} g^{\rho_{1}} \right) b^{\rho_{3}} \tilde{\cal L}^{...} = 0.
\label{fgb}
\ee

Applying
$
Z^{*}
$
one finds out that also

\be
\sum_{(\rho_{1},\rho_{2},\rho_{3})} 
\left( f^{\rho_{1}} g^{\rho_{2}} + 
f^{\rho_{2}} g^{\rho_{1}} \right) c^{\rho_{3}} \tilde{\cal L}^{...} = 0.
\label{fgc}
\ee

In particular these relations make the starting point -relation
(\ref{II2-compact})- identically satisfied. Because the indices
$A, B, D$
in the relation (\ref{II2-Fock}) are arbitrary we have proved in fact that this
relation is equivalent to the following two relations:

\be
\sum_{(\rho_{1},\rho_{2},\rho_{3})} 
\left[ b^{(\rho_{1})}_{(B)} c^{(\rho_{2})}_{(B)} + 
b^{(\rho_{2})}_{(B)} c^{(\rho_{1})}_{(B)} \right] b^{(\rho_{3})}_{(A)} 
\tilde{\cal L}^{...} = 0
\label{II-final1}
\ee

and

\be
\sum_{(\rho_{1},\rho_{2},\rho_{3})} 
\left[ b^{(\rho_{1})}_{(B)} c^{(\rho_{2})}_{(B)} + 
b^{(\rho_{2})}_{(B)} c^{(\rho_{1})}_{(B)} \right] c^{(\rho_{3})}_{(A)} 
\tilde{\cal L}^{...} = 0
\label{II-final2}
\ee

for any
$
A \not= B.
$

It is easy to see that the preceding two relations make (\ref{II1-Fock}) and 
(\ref{II2-Fock}) identities.

Now, starting for instance from (\ref{II-final1}) one can use recurrsively 
the redefinition trick from the preceding Section to show the 
$
\tilde{\cal L}^{...}
$
can be changed such that it verifies:

\be
\left[ b^{(\mu)}_{(1)} b^{(\nu)}_{(B)} + 
b^{(\nu)}_{(1)} b^{(\mu)}_{(B)} \right] \tilde{\cal L}^{...} = 0
\ee

for
$
B = 2,3,...,N.
$

The preceding relation immediately implies the relation (\ref{antisymmetry})
and accordingly theorem \ref{structureN}.
This finishes the proof of the main theorem.

\section{Main Theorem}

6.1 In this Section we will exhibit the result of theorem \ref{structureN} in a
more compact way. To do this we introduce the second-order hyper-Jacobians 
in full generality:

\be
J^{A_{1},...,A_{k};\mu_{k+1},...,\mu_{n}}_{\nu_{1},...,\nu_{k}} \equiv
\varepsilon^{\mu_{1},...,\mu_{n}} \prod_{i=1}^{k} 
\psi^{A_{i}}_{\mu_{i}\nu_{i}} \quad (k = 0,...,n).
\label{hJ}
\ee

One notices that these expressions are the natural generalization of the
expressions (\ref{hyperJ}) defined in Section 3 and that they have
properties of the same kind; namely a antisymmetry property:

\be
J^{A_{P(1)},...,A_{P(k)};\mu_{Q(k+1)},...,\mu_{Q(n)}}_{\nu_{P(1)},...,
\nu_{P(k)}} = (-1)^{|P|+|Q|}
J^{A_{1},...,A_{k};\mu_{k+1},...,\mu_{n}}_{\nu_{1},...,\nu_{k}}  
\quad (k = 0,...,n)
\label{antisym-hJ}
\ee
(where
$P$ 
is a permutation of the numbers
$1,...,k$
and 
$Q$
is a permutations of the numbers 
$k+1,...,n$)
and a tracelessness property (see \cite{O}):

\be
J^{A_{1},...,A_{k};\mu_{k+1},...,\mu_{n-1},\zeta}_{\nu_{1},...,
\nu_{r-1},\zeta} = 0.
\quad (k = 1,...,n-1).
\label{trace-hJ}
\ee

The relations (\ref{antisym}) and (\ref{trace}) are particular cases of these
relations (for the case of the scalar field). 

Then we have

\begin{thm} 

The most general solution of the equations (\ref{II1}) and (\ref{II2}) is of
the following form:

\be
{\cal L} = \sum_{k=0}^{n} 
{\cal L}^{\nu_{1},...,\nu_{k}}_{A_{1},...,A_{k};\mu_{k+1},...,\mu_{n}}
J^{A_{1},...,A_{k};\mu_{k+1},...,\mu_{n}}_{\nu_{1},...,\nu_{k}} 
\label{polyn-hJ}
\ee
where the functions
$
{\cal L}^{...}_{...}
$
are independent of
$
\psi_{\mu\nu}^{B}
$:

\be
\partial^{\mu\nu}_{B}
{\cal L}^{\nu_{1},...,\nu_{k}}_{A_{1},...,A_{k};\mu_{k+1},...,\mu_{n}} = 0
\quad (k = 0,...,n),
\ee
and have analogous properties as the hyper-Jacobians, namely the
(anti)symmetry property:

\be
{\cal L}_{A_{P(1)},...,A_{P(k)};\mu_{Q(k+1)},...,\mu_{Q(n)}}^{\nu_{P(1)},...,
\nu_{P(k)}} = (-1)^{|P|+|Q|}
{\cal L}_{A_{1},...,A_{k};\mu_{k+1},...,\mu_{n}}^{\nu_{1},...,\nu_{k}}  
\quad (k = 0,...,n)
\label{antisym-l-hJ}
\ee
(where
$P$ 
is a permutation of the numbers
$1,...,k$
and 
$Q$
is a permutations of the numbers 
$k+1,...,n$) 
and also verify the identities:

\be
{\cal L}_{A_{1},...,A_{k};\mu_{k+1},...,\mu_{n-1},\zeta}^{\nu_{1},...,
\nu_{k-1},\zeta} = 0.
\label{trace-l-hJ}
\ee
(i. e. are traceless). The function coefficients
$
{\cal L}^{...}_{...}
$
are uniquely determined by
$
{\cal L}
$
and the properties (\ref{antisym-l-hJ}) and (\ref{trace-l-hJ}) above.
\label{structure-hJ}
\end{thm}

{\bf Proof:} For
$
N = 1
$
this result coincides with theorem \ref{structure}. For
$
N \geq 2
$
we will prove it using the results from the preceding two sections. Namely, we
will show that the expression (\ref{polynN-compact-dual}) can be rearranged 
such that it coincides with (\ref{polyn-hJ}) above. In fact, it is easier to 
start with (\ref{polyn-hJ}) and to obtain the previous expression
(\ref{polynN-compact-dual}). To this purpose we will make
$
N \rightarrow N+1
$
and suppose that the indices 
$
A, B, ...
$
run from 
$0$
to
$N$;
the indices
$
a, b, ...
$
will run from
$1$
to 
$N$.

We separate in the  expression (\ref{polyn-hJ}) the contributions in which 
$
n - r
$
indices take values from
$1$
to 
$N$
and
the rest take the value
$0$.

One obtains

\be
{\cal L} = \sum_{k=0}^{n} \sum_{r=k}^{n} C^{n-r}_{n-k}
{\cal L}^{\mu_{k+1},...,\mu_{n}}_{0,...,0,a_{r+1},...,a_{k};
\nu_{1},...,\nu_{k}}
J^{0,...,0,a_{r+1},...,a_{k};\nu_{1},...,\mu_{k}}_{\mu_{k+1},...,\mu_{n}} 
\label{polyn-doublesum}
\ee
where it is understood that there are 
$
r - k
$
entries equal to
$0$.

One can rearrange this expression as follows:

\be
{\cal L} = \sum_{k=0}^{n} 
l^{\mu_{k+1},...,\mu_{n}}_{a_{k+1},...,a_{k};\nu_{1},...,\nu_{k}}
J^{a_{k+1},...,a_{n};\nu_{1},...,\nu_{k}}_{\mu_{k+1},...,\mu_{n}} 
\label{polyn-a}
\ee

where

\be
l^{\mu_{k+1},...,\nu_{n}}_{a_{k+1},...,a_{n};\nu_{1},...,\nu_{k}} \equiv
\sum_{r=0}^{k} C^{n-r}_{n-k} 
{\cal A}_{\nu_{1},...,\nu_{k}} 
\left({\cal L}^{\mu_{r+1},...,\mu_{n}}_{0,...,0,a_{k+1},...,a_{k};
\nu_{1},...,\nu_{r}} \prod_{l=r+1}^{k} \psi^{0}_{\mu_{l}\nu_{l}} \right)
\ee
where
$
{\cal A}_{\nu_{1},,...,\nu_{k}}
$
is the projector which antisymmetrizes in the indices
$
\nu_{1},...,\nu_{k}
$
and there are
$k-r$
entries equal to
$0$.

One defines the dual tensor
$
\tilde{l}^{...}_{...}
$
by analogy to (\ref{dual}) and discovers after some combinatorics that it is
given by the following relation:

\be
\tilde{l}^{\mu_{k+1},...,\nu_{n};\nu_{k+1},...,\nu_{n}}_{a_{k+1},...,a_{n}} 
= \sum_{s=0}^{k} 
L^{\mu_{k+1},...,\nu_{n};\nu_{k+1},...,\nu_{n};
\sigma_{1},...,\sigma_{s}}_{a_{k+1},...,a_{n};\rho_{s+1},...,\rho_{n}}
J^{(0)\rho_{s+1},...,\rho_{n}}_{\sigma_{1},...,\sigma_{s}}
\ee

where

\be
L^{\mu_{k+1},...,\mu_{n};\nu_{k+1},...,\nu_{n};
\sigma_{1},...,\sigma_{s}}_{a_{k+1},...,a_{n};\rho_{s+1},...,\rho_{n}} \equiv
{\rm const.} {\cal A}_{\rho_{s+1},...,\rho_{n}} \left(
{\cal L}^{\mu_{k+1},...,\nu_{n};\nu_{k+1},...,\nu_{n};
\sigma_{1},...,\sigma_{s},\mu_{k+1},...,\mu_{n}}_{0,...,0,
a_{k+1},...,a_{n};\rho_{s+1},...,\rho_{k}} 
\prod_{l=k+1}^{n} \delta^{\nu_{l}}_{\rho_{l}} \right)
\ee
and we have
$s$
entries equal to
$0$.

If one defines the dual tensor
$
\tilde{L}^{....}  
$
as in (\ref{dual}), one finally finds out that

\be
L^{\mu_{k+1},...,\nu_{n};\nu_{k+1},...,\nu_{n};
\sigma_{1},...,\sigma_{s}}_{a_{k+1},...,a_{n};\rho_{s+1},...,\rho_{n}} =
{\rm const} 
\tilde{\cal L}^{\sigma_{1},...,\sigma_{s}\mu_{k+1},...,\mu_{n};
\rho_{1},...,\rho_{s}\nu_{k+1},...,\nu_{n}}_{0,...,0,a_{k+1},...,a_{n}}
\ee
and we have 
$s$
entries equal to 
$0$.

So, finally, after some relabelling we get the following expression from
(\ref{polyn-hJ}):

\be
{\cal L} = \sum_{k=0}^{n} \sum_{r=0}^{n-k} {\rm const.}
\tilde{\cal L} ^{\sigma_{1},...,\sigma_{r}\mu_{1},...,\mu_{k};
\rho_{1},...,\rho_{r},\nu_{1},...,\nu_{k}}_{0,...,0,a_{1},...,a_{k}}
\tilde{J}^{a_{1},...,a_{k}}_{\mu_{1},...,\mu_{k};\nu_{1},...,\nu_{k}}
\tilde{J}^{(0)}_{\rho_{1},...,\rho_{r};\sigma_{1},...,\sigma_{r}}
\ee

It is clear that we can iterate the procedure with the indices
$1,...,N$
and we will obtain finally the expression (\ref{polynN-compact-dual}).
$\qed$

6.2 It is interesting to define the so-called {\it horizontalisation} 
operation 
${\bf h}$
on the space on differential form on the jet-bundle space (see for
instance \cite{K4}). In particular, it is defined by linearity, 
multiplicity and:

\be
{\bf h} dx^{\mu} \equiv dx^{\mu}, \quad 
{\bf h} d \psi^{A} \equiv \psi^{A}_{\mu} dx^{\mu}, \quad 
{\bf h} d \psi^{A}_{\mu} \equiv \psi^{A}_{\mu\nu} d x^{\nu}, \quad
{\bf h} d \psi^{A}_{\mu\nu} \equiv \psi^{A}_{\mu\nu\rho} d x^{\rho}.
\ee

Let us now define the differential form

\be
\Lambda \equiv \sum_{k=0}^{n} (-1)^{k(n+1)} 
{\cal L}^{\mu_{k+1},...,\mu_{n}}_{A_{k+1},...,A_{n};\nu_{1},...,\nu_{k}}
d \psi^{A_{k+1}}_{\mu_{k+1}} \wedge \cdots \wedge d \psi^{A_{n}}_{\mu_{n}}
\wedge d x^{\nu_{1}} \wedge \cdots d x^{\nu_{k}}.
\ee

Then it is elementary to prove 

\begin{prop}

The Euler-Lagrange form associated to the trivial Lagrangian (\ref{polyn-hJ})
is given by

\be
L = {\bf h} \Lambda.
\label{hor}
\ee
\end{prop}

In other words, any second-order trivial Lagrangian can be obtained as a 
result of the horizontalisation operation. 

\begin{rem}

It is elementary to prove the same assertion for the first-order trivial
Lagrangians. In fact, relation (\ref{hor}) stands true if instead of the form
$\Lambda$
above one uses the expression (\ref{lam}).
\end{rem}

6.3 We still did not exploited the other two conditions of triviality, namely
(\ref{II3}) and (\ref{II4}). This analysis of these relations proves to be 
not very conclusive. Indeed it is very hard to obtain some relations involving
directly the coefficients
$
{\cal L}^{...}_{...}
$
from (\ref{polyn-hJ}). The best we could do was to re-express this formula as

\be
{\cal L} = \sum_{k=0}^{n} {1 \over k!} 
L^{\mu_{1},...,\mu_{k};\nu_{1},...,\nu_{k}}_{A_{1},...,A_{k}}
\prod_{i=1}^{k} \psi^{A_{i}}_{\mu_{i}\nu_{i}}
\ee

where the functions
$
L^{...}_{...}
$
are completely symmetric in the couples
$
(A_{l},\mu_{l},\nu_{l}) 
$
and are invariant at the interchanges
$
\mu_{l} \leftrightarrow \nu_{l} \quad (l = 0,...,n).
$

Then one can obtain from (\ref{II3}) and (\ref{II4}) some rather 
complicated differential equations on the functions
$
L^{...}_{...}.
$
These equations seems to be rather hard to analyse. In particular it is not
clear if a polynomial structure in the first-order derivatives can be
established, although this seems not very plausible.

\section{Final Comments}

It is quite reasonable to expect that in the same way one can prove a formula
of the type (\ref{polynN-compact}) for the case of a trivial Lagrangian
of arbitrary order. In the 
$p$-th
order case one must have hyper-Jacobians (\ref{dual-hyperJN}) with 
$p$
groups of indices of the type
$
\rho_{1},...,\rho_{r}.
$

One must also expect that a trivial Lagrangian, more precisely its 
dependence on the highest order derivatives, is also a result of the
horizontalisation operation.

These problems deserve some investigation and we intend to do this in future.
\vskip 1cm

%{\bf Acknowledgments:} 

%\newpage

\end{document}